\documentclass[twocolumn,pra,superscriptaddress,aps,showpacs,showkeys,amsmath,floatfix]{revtex4-1}

\usepackage{graphics}
\usepackage{graphicx}
\usepackage{epstopdf}
\usepackage{dcolumn}
\usepackage{bm}
\usepackage{longtable}
\usepackage{epsfig}
\usepackage{times}
\usepackage{url}
\usepackage{color}

\begin{document}
\title{Generation of short hard X-ray pulses of tailored duration using a M\"ossbauer source}

\author{Guan-Ying   \surname{Wang}}
\affiliation{Department of Physics, National Central University, Taoyuan City 32001, Taiwan}

\author{Wen-Te \surname{Liao}}
\email{wente.liao@g.ncu.edu.tw}
\affiliation{Department of Physics, National Central University, Taoyuan City 32001, Taiwan}
\affiliation{Max Planck Institute for Nuclear Physics, Saupfercheckweg 1, 69117 Heidelberg, Germany}
\date{\today}
\begin{abstract}
We theoretically investigate a scheme for generations of single hard X-ray pulses of controllable duration in the range of 1 ns - 100 ns from a radioactive M\"ossbauer source.
The scheme uses a magnetically perturbed $^{57}$FeBO$_3$ crystal illuminated with recoilless 14.4 keV photons from a radioisotope $^{57}$Co nuclide. 
Such compact X-ray source is useful for the extension of quantum optics to 10 keV energy scale which has been spotlighted in recent years. So far, experimental achievements are mostly performed in synchrotron radiation facilities. However, tabletop and portable hard X-ray sources are still limited for time-resolved measurements  and for implementing coherent controls over nuclear quantum optics systems. 
The availability of compact hard X-ray sources may become the engine to apply  schemes of quantum information down to the subatomic scale.
We demonstrate that the present method is versatile and provides an economic solution utilizing a M\"ossbauer source to perform time-resolved nuclear scattering, to produce suitable pulses for photon storage and to flexibly generate X-ray single-photon entanglement. 
\end{abstract}

\maketitle
X-ray quantum optics \cite{adams2013} opens the completely new angstrom wavelength regime   for studying light-matter interactions. Recently  many remarkable experiments \cite{Shvydko1996, Buth2007, Tamasaku2007, Glover2009, Roehlsberger2010, Rohringer2012, Roehlsberger2012, Glover2012, Shwartz2012, Heeg2013, Heeg2015a, Heeg2015b, Heeg2017} have been achieved with synchrotron radiation (SR) or X-ray free electron lasers (XFELs).
Many new systems \cite{Buervenich2006, Palffy2009, Shwartz2011, Liao2011, Liao2012a, Liao2013, Liao2014, Liao2016, Kong2016, Cavaletto2014, Gunst2016l, Gunst2016e, Olga1999, Liao2017} based on X-ray facilities were also theoretically investigated. 
One of reasons that SR becomes superior to a M\"ossbauer source namely radioisotope is the very short time structure of SR.
When utilizing SR to excite nuclear transitions, its picosecond pulse duration allows for the discrimination of resonantly scattered photons from both background radiation of SR and instantaneous nonresonant charge scattering in time domain \cite{Roehlsberger2004, Shenoy2007}. 
This method, remarkably  suggested by S. L. Ruby  \cite{Shenoy2007}, made great success in the field of nuclear condensed matter physics \cite{Roehlsberger2004} and lays the basis of many nuclear schemes of  X-ray quantum optics \cite{Shvydko1996, Roehlsberger2010, Roehlsberger2012, Heeg2013, Heeg2015a, Heeg2015b, Heeg2017, Palffy2009, Liao2011, Liao2013}.
It is therefore of great interest to find methods of producing short hard X-ray pulses of suitable temporal profile from typical M\"ossbauer sources \cite{Olga2014} since  a radioisotope is portable and more affordable than a X-ray facility.
Moreover, for many  applications like quantum information \cite{Palffy2009, Olga2014, Kong2016, Gunst2016l, Gunst2016e} it is more convenient to use a compact X-ray source like radioisotope than a SR facility. 
The unperturbed time structure of emitted X-ray pulses from a typical M\"ossbauer source, however, may not be suitable for different purposes. In a photon storage scheme \cite{Kong2016} for example, the moderate duration of a pulse to be stored is preferable, i.e.,  shorter than the coherence time of a memory medium but longer than the inverse of the memory bandwidth. Generation of such single hard X-ray pulses is not trivial with available schemes \cite{Kong2016, Olga2014}.
%
%
It is therefore important to develop methods to  generate X-ray wave packets of mission-oriented duration from a conventional M\"ossbauer source   \cite{Liao2012a, Olga2014, Kong2016}.
We here investigate an all-field scheme using a magnetically perturbed $^{57}$FeBO$_3$ crystal \cite{Smirnov1984, Shvydko1993, Shvydko1996} to generate short single hard X-ray pulses of adjustable duration in the 1 ns - 100 ns range.
The present tabletop scheme can be useful for time-resolved measurements, X-ray quantum optics and even probing the gravitational deflection of X-rays in any laboratory \cite{Liao2015}.
\begin{figure}[b]
\vspace{-0.4cm}
  \includegraphics[width=0.45\textwidth]{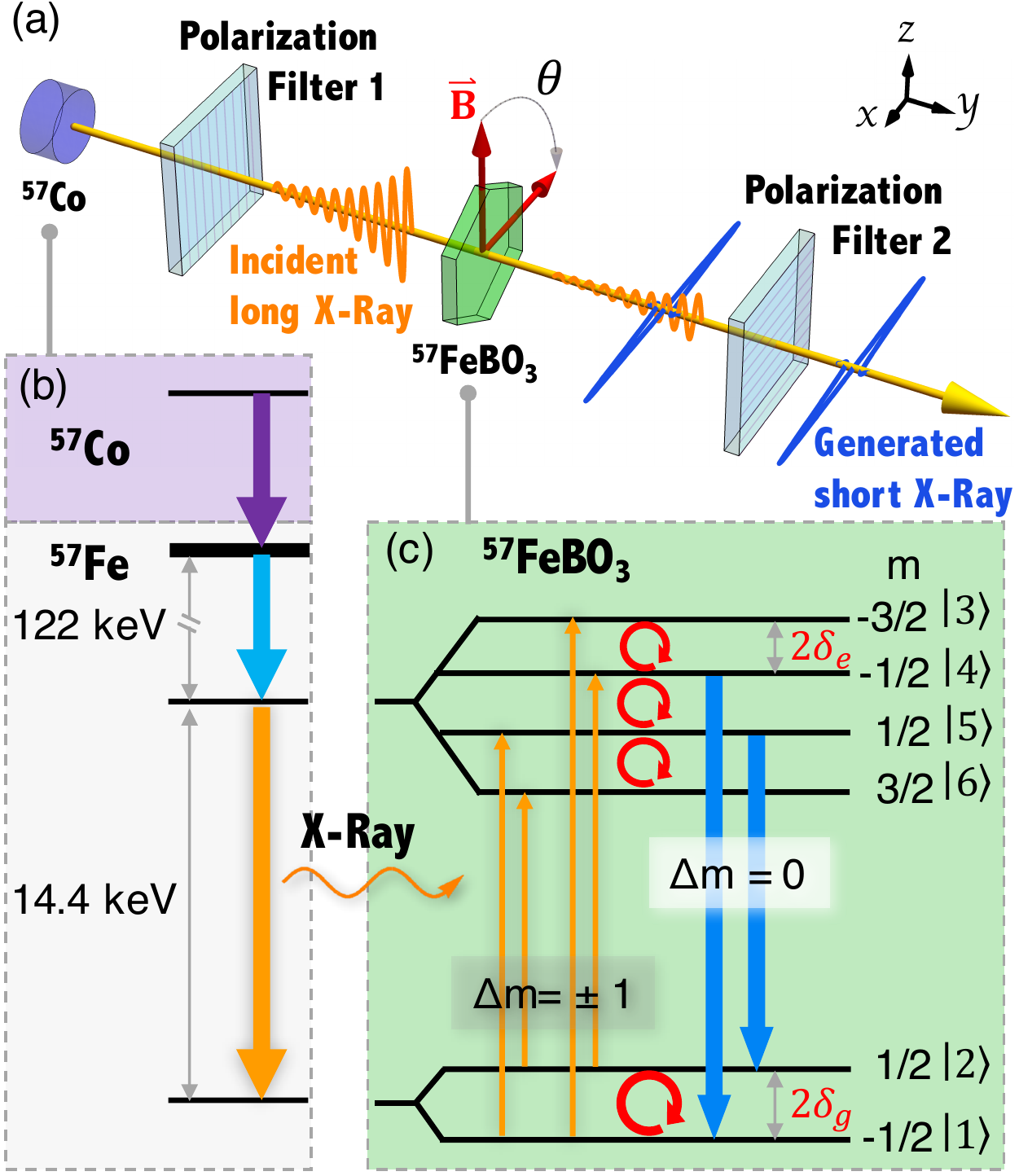}
  \caption{\label{fig1}
(a) a vertically (along z-axis) polarized X-ray of 14.4 keV (orange wavy curve) from a $^{57}$Co M\"ossbauer source impinges on a $^{57}$FeBO$_3$ crystal (green hexagon). A horizontally (along x-axis) polarized X-ray (blue sharp curve) is generated when properly switching the magnetization of a $^{57}$FeBO$_3$ by an external magnetic field $\vec{B}$ (red rotating arrows).
(b) in a radioisotope source, the $^{57}$Co nuclide becomes $^{57}$Fe via electron capture, and then the decayed $^{57}$Fe nucleus emits $\gamma$-rays of 122 keV and subsequently  hard X-rays of 14.4 keV polarized by polarization filter 1. 
(c) 
the vertically polarized X-rays drive $\Delta m=\pm 1$ nuclear transitions of $^{57}$Fe. By dynamically switching the angle $\theta$ between the z-axis and $\vec{B}$, the hyperfine splitting $2\delta_g$ ($2\delta_e$) among ground (excited) states are modulated. 
Due to the magnetic switching, the emission of the X-rays with horizontal polarization via two $\Delta m=0$ transitions is controllable.
  }
\end{figure}

%
Figure~\ref{fig1} illustrates the setup. In a M\"ossbauer source the radioactive $^{57}$Co nuclei become excited $^{57}$Fe nuclei via electron capture. The excited $^{57}$Fe nuclei then experience a cascade decay and emit in turn 122 keV and 14.4 keV photons. In a typical M\"ossbauer experiment the detection of former triggers the time sequence of probing a nuclear target of interest using latter \cite{Shvydko1993, Olga2014}.
A recoilless and vertically polarized \cite{Shvydko1993, Szymanski2006} 14.4 keV X-ray, of duration 141 ns, from a radioisotope $^{57}$Co source (see Fig.~\ref{fig1}(b)) impinges on an isotopically enriched $^{57}$FeBO$_3$ crystal. 
We invoke the elegant magnetic property of  an  iron borate $^{57}$FeBO$_3$ crystal whose strong  magnetization can be abruptly rotated via an external weak magnetic field $\vec{B}$ of few Gauss in $<$ 1 ns \cite{Kolotov1998}. 
A recent study with femtosecond laser pulses suggests an even faster control over iron borate \cite{Afanasiev2014}.
This property allows for fast switching of the internal magnetic hyperfine field at the  $^{57}$Fe nuclei, and therefore renders controls  over the dynamics of nuclear excitations possible \cite{Shvydko1993,Shvydko1996,Palffy2009}. 
The $^{57}$Fe nuclear level scheme with hyperfine splitting of the ground and excited states of spins respectively $I_g = 1/2$ and $I_e = 3/2$ is illustrated in Fig.~\ref{fig1}(c).
When choosing the z-axis as the  quantization axis, the vertically  polarized X-ray (VPX)  is treated as a superposition of right circularly polarized X-rays (RCPX)  and left circularly polarized X-rays (LCPX).
Therefore,  VPX drives $\Delta m=\pm 1$ nuclear  magnetic dipole  transitions of $^{57}$Fe in the crystal.
The projections of the crystal magnetization in the quantization axis causes 
hyperfine splitting  proportional to $\cos\theta$ and
simultaneously
Larmor precession about the x-axis with frequencies proportional to $\sin\theta$. 
The Larmor precession leads  to a crosstalk between orthogonal polarizations of a single photon. One can therefore  switch the crystal magnetization back and forth to create a time window allowing for a transient crosstalk, which transforms a temporal slice of a long pulse of a specific polarization into another.
In what follows, we will invoke the above crosstalk dynamics to control the duration of  produced X-ray pulses.

\begin{figure}[b]
\vspace{-0.4cm}
  \includegraphics[width=0.45\textwidth]{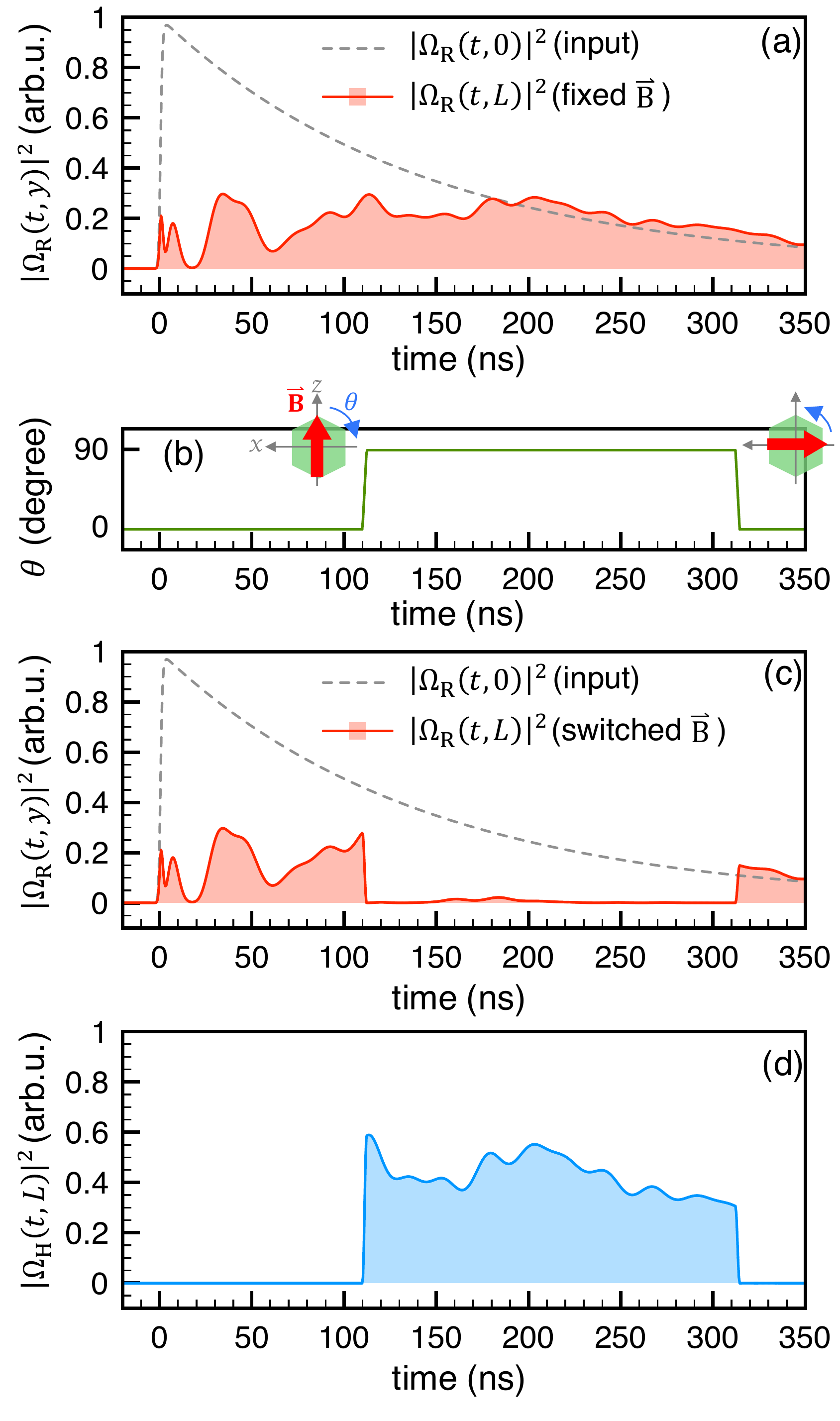}
  \caption{\label{fig2}
(a) without magnetic switching (fixed $\vec{B}$)
red solid line depicts the intensity of the scattered right circularly polarized X-ray $\vert \Omega_R\left( t,L\right) \vert^2$. The gray dashed line depicts the input $\vert \Omega_R\left( t,0\right) \vert^2$.
With the magnetic switching whose time sequence  is demonstrated in panel  (b),
the magnetic rotation transfers a temporal fraction of (c) right circularly polarized X-ray  to (d) horizontally polarized component. 
Two cartoons in panel  (b) visualize 90$^{\circ}$ magnetic rotations applied to a $^{57}$FeBO$_3$ crystal at around 110 ns and 310 ns. Red arrows represent the applied magnetic field, blue curved arrows show the switching paths, and green hexagons illustrates the $^{57}$FeBO$_3$ crystal.
}
\end{figure}
\begin{figure}[hb]
\vspace{-0.4cm}
  \includegraphics[width=0.45\textwidth]{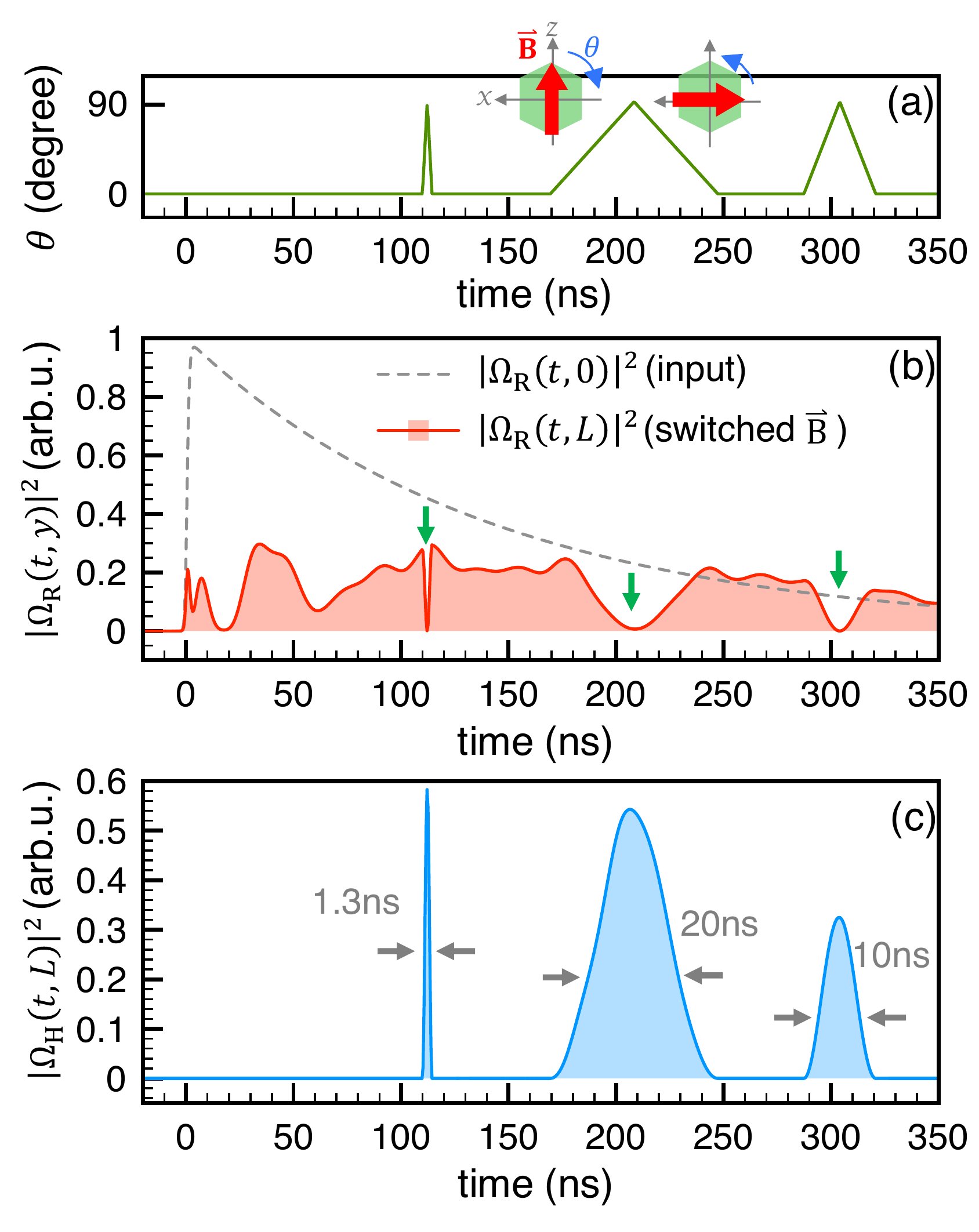}
  \caption{\label{fig3}
Panel (a) depicts three magnetic rotations at different rates applied to a $^{57}$FeBO$_3$ crystal. All single pulses are generated by two opposite 90$^{\circ}$ rotations of $\vec{B}$. Two cartoons visualize the 90$^{\circ}$ magnetic rotations during 170 ns and 250 ns.
Panel (b) illustrates the input (gray dashed line) and output (red solid line) right circularly polarized X-rays. Three vertical green arrows indicate three temporal splits cut by magnetic switching at different rates. 
Panel (c) demonstrates the generation of three horizontally polarized single X-ray pulses with durations of 1.3 ns, 20 ns and 10 ns. 
}
\end{figure}
%
%
Figure~\ref{fig2} demonstrates the effects of magnetic switching whose dynamics is described by Optical-Bloch equation (OBE)  \cite{Liao2012a}.
It is convenient to fix the quantization axis along the z-axis and then rotate the hyperfine splitting Hamiltonian at each $\theta$  back onto the z-axis.
The OBE, which is numerically solved  by using Runge Kutta 4th order method, reads
\begin{eqnarray}\label{eq1}
 &  & \partial_{t}\hat{\rho}=\frac{1}{i\hbar}\left[\hat{H},\hat{\rho}\right]+\mathcal{L}\rho,\nonumber\\
 &  & \hat{H} = \hat{H}_i + \hat{H}^z_h +\frac{d\theta}{dt}\hat{J}_y ,\nonumber\\ 
 &  & \frac{1}{c}\partial_{t}\Omega_{R}+\partial_{y} \Omega_{R}=i\eta\left( a_{51}\rho_{51}+a_{62}\rho_{62}\right) ,\nonumber\\
 &  & \frac{1}{c}\partial_{t}\Omega_{H}+\partial_{y} \Omega_{H}=i\eta\left( a_{41}\rho_{41}+a_{52}\rho_{52}\right) ,\nonumber\\
 &  & \frac{1}{c}\partial_{t}\Omega_{L}+\partial_{y} \Omega_{L}=i\eta\left( a_{31}\rho_{31}+a_{42}\rho_{42}\right) ,
\end{eqnarray}
where $\eta=6 \Gamma\alpha/L$, $\alpha$ is the resonant thickness, $\Gamma =$1/141 GHz the radioactive decay rate of nuclear excited state,  $L$ the crystal thickness.
$\hat{\rho}$ is the density matrix for the state vector
$\sum_{i=1}^{6}A_{i}\vert i\rangle$ of the 6-level $^{57}$Fe nuclei, and each coherence $\rho_{ij}=A_i A_j^*$. 
$\hat{H}_i$ is the interaction Hamiltonian
describing the nucleus-X-ray coupling. $\hat{H}_h^z$ depicts the rotated hyperfine Hamiltonian.
$a_{ij}$
are Clebsch-Gordan coefficients of corresponding transitions. 
$\hat{J}_y$ is the y-component of the angular momentum operator when  magnetic field is switched about the y-axis (see Appendix B) and
$\mathcal{L}\rho$ describes the radioactive decay of the excited states.
$\Omega_{R}\left( t, y\right) $, $\Omega_{L}\left( t, y\right)$ and $\Omega_{H}\left( t, y\right)$ are Rabi frequencies of RCPX, LCPX and the horizontally polarized X-ray (HPX) (parallel to the x-axis in Fig.~\ref{fig1}(a)), respectively.
Each wave equation describes the forward propagation of X-rays of different polarization.
The explicit form of all Hamiltonian matrixes is given in  Appendix A.
A comparison will be drawn between the case with a fixed  $\vec{B}$ and that with a switched $\vec{B}$. 
The gray dashed lines in Fig.~\ref{fig2} show the input X-ray intensity from the $^{57}$Co nuclide.
For both cases the inputs are $\Omega_{R}\left( t,0\right)=\Omega_{L}\left( t,0\right)=e^{-\Gamma t/2}$.
The used $^{57}$FeBO$_3$ parameters are $\alpha=30$, hyperfine splitting of ground states $\delta_g=12.63\Gamma$ and that of excited states $\delta_e=7.37\Gamma$.
We first show the output X-ray intensity for a fixed $\vec{B}$ in Fig.~\ref{fig2}(a) where the input and output polarizations are the same (termed as the unperturbed signal).
Two identical output RCPX $\vert \Omega_R\left( t,L\right) \vert^2$ (red solid line) and LCPX  $\vert \Omega_L\left( t,L\right) \vert^2$ (not shown) results from the interference between nuclear currents due to $\Delta m = 1$ and $\Delta m = -1$ transitions, respectively.
Fig.~\ref{fig2}(b), (c) and (d) illustrate an example in which we continuously switch the external $\vec{B}$. 
The time sequence of the rotated magnetic field is given by Fig.~\ref{fig2}(b). 
Two continuous but opposite 90$^{\circ}$ magnetic switchings are applied to a $^{57}$FeBO$_3$ crystal, and their switching paths are depicted by two cartoons. 
The first clockwise switching from 0$^{\circ}$ to 90$^{\circ}$ is implemented at around 110 ns, and then the $\vec{B}$ field remains at 90$^{\circ}$ for 200 ns. 
Subsequently the second counter-clockwise rotation switches the magnetization back to its initial orientation at around 310 ns. 
As demonstrated in Fig.~\ref{fig2}(c) and (d), the above magnetic switching transfers a temporal fraction of  RCPX, namely, the missing part during 110 ns $< t <$ 310 ns in Fig.~\ref{fig2}(c),
to HPX. Because both RCPX and LCPX contribute equal amounts of energy to HPX, the HPX intensity is the double of the transferred part in unperturbed RCPX.

Given the capability of transferring photons from one polarization to another by the above scheme, 
in Fig.~\ref{fig3} we illustrate how to use the method to generate short hard X-ray pulses of tailored pulse durations. 
Fig.~\ref{fig3}(a) depicts a sequence of switching $\vec{B}$ back and forth between 0$^{\circ}$ and 90$^{\circ}$ 
at three different rates. As one can see in Fig.~\ref{fig3}(b), the magnetic switching cuts three splits 
with corresponding width in the perturbed RCPX signal (indicated by vertical green arrows). 
Our calculation shows that an identical situation also happens to LCPX. 
Fig.~\ref{fig3}(c) illustrates that the missing fractions of circularly polarized X-rays are transferred to HPX as individual pulses having durations of, in turn, 
1.3 ns, 10 ns and 20 ns depending on how fast  $\vec{B}$ is rotated. 
With a given resonant thickness and hyperfine splitting of a $^{57}$FeBO$_3$ crystal,
the peak intensity and peak position of a generated single pulse can  be adjusted by the switching process. For example, in Fig.~\ref{fig3}(c), the reason that the shortest pulse has the highest intensity is because the switched $\vec{B}$ reaches the angle of 90$^{\circ}$ at the instant when the unperturbed $\vert \Omega_R\left( t,L\right)\vert^2$ is maximum. The slight asymmetry of each pulse represents the ripples of the unperturbed circularly polarized X-rays.
Also, the speed of switching controls the  generated pulse durations, which highlights the flexibility of the method.

We estimate the photon counting rates in what follows.
For a 200 mCi  commercial M\"ossbauer source, forward emission in a solid angle of $5 \times 10^{-4}$ sr and the analysis of Clebsch-Gordan coefficients, 
the VPX photon counting rate is about 96300/s at the first $^{57}$FeBO$_3$ crystal. 
The produced HPX counting rate depends on the generated pulse area. For single pulses in  Fig.~\ref{fig3}(c), the rates are in turn estimated to be 430/s, 6600/s and 1800/s. According to our calculations, it is possible to triple these rates by using a thicker crystal of $\alpha=100$, $\delta_e=31\Gamma$ and $\delta_g=53\Gamma$ (see Appendix C). 
Given the VPX counting rate of 90 kHz,  the switching may repeat every 0.01 millisecond which should cause much less heating (by $\ll$ 15 Kelvin) of the crystal than MHz repetitation rate \cite{Shvydko1993}.
Also, during switching, the  magnetoelastic oscillations with frequency of  few MHz  will cause modulations on longer X-ray pulses  \cite{Smirnov1984,Shvydko1993,Kolotov1998}. 
However, the duration of the polarization transfer is mainly controlled by quantum dynamics.
Solutions to suppress the intensity of 
magnetoelastic oscillations have been demonstrated, e.g., using several thin crystals instead of a thick one \cite{Smirnov1984}.

\begin{figure}[t]
\vspace{-0.4cm}
  \includegraphics[width=0.43\textwidth]{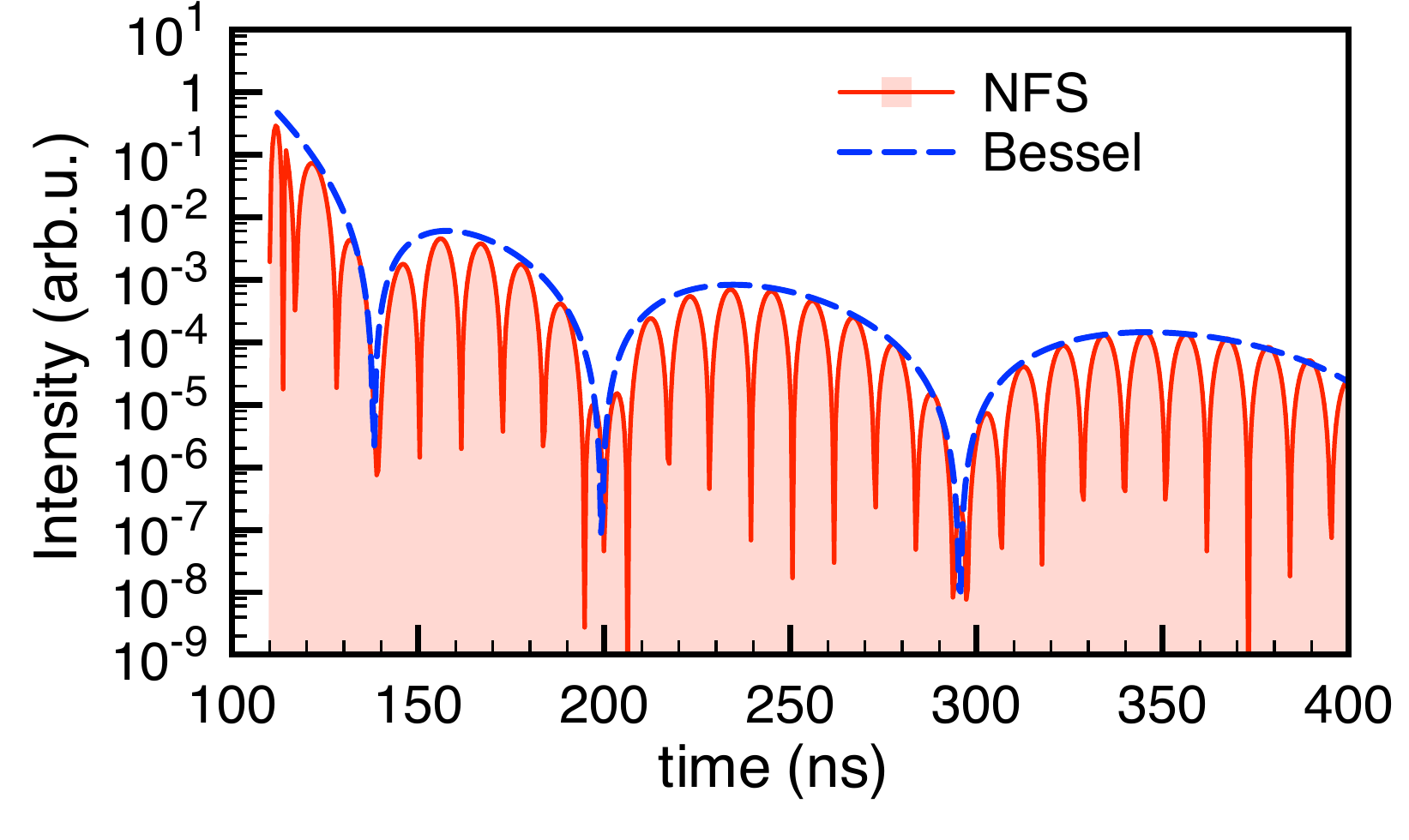}
  \caption{\label{fig4}
The generated leftmost single pulse of duration of 1.3 ns in Fig.~\ref{fig3}(c) is used to probe a downstream nuclear target with nuclear forward scattering. Red solid line illustrates the NFS signal and blue dashed line shows the typical Bessel fitting.
}
\end{figure}

We point out that the present scheme is a solution for some potential applications which are so far considered only suitable for SR facilities:
(1) Fig.~\ref{fig2} and Fig.~\ref{fig3} in fact demonstrate flexible generations of X-ray single-photon entanglement in polarization \cite{Palffy2009, Shwartz2011}, namely, the single-photon wave packet (Fig.~\ref{fig2}(c,d) and Fig.~\ref{fig3}(b,c)) is described by $\vert\psi\rangle = \vert 1\rangle_V \vert 0\rangle_H+\vert 0\rangle_V \vert 1\rangle_H$. Our calculation shows that a successful production of entanglement via the present scheme is not sensitive to switching moment \cite{Palffy2009}. It therefore allows for freely choosing when to turn on the polarization transducer as demonstrated in Fig.~\ref{fig2} and Fig.~\ref{fig3}.
(2) tailored pulse duration for versatile use. It is not trivial to produce an X-ray pulse with a mission-oriented duration, e.g., 28 ns for X-ray photon storage \cite{Kong2016}. As depicted in Fig.~\ref{fig3}, the present scheme offers a solution for such purpose by changing the switching rate. Based on our calculation, an hard X-ray pulse of a duration in the range of 1 ns - 100 ns can be produced by the present scheme.
(3) a possible use for time-resolved measurements.
We utilize the produced HPX of the duration of 1.3 ns, i.e., the leftmost peak in Fig. \ref{fig3}(c), as the incident field for the second $^{57}$Fe-enriched target to be probed by nuclear forward scattering (NFS).
Fig.~\ref{fig4} depicts the scattered X-ray of the second target whose parameters are resonant depth $\alpha_2=20$, $\delta_g=25.26\Gamma$ and  $\delta_e=14.74\Gamma$.
The envelope (blue dashed  line) of the NFS signal is so called dynamical beat \cite{Shvyd1999H} and is described by $\left( \frac{\alpha_2}{\sqrt{\alpha_2\Gamma t}}J_{1}\left[2\sqrt{\alpha_2\Gamma t}\right]\right)^{2} e^{-\Gamma t}$ where  $J_{1}$ is the Bessel function of the first kind \cite{Shvyd1999H}.
The modulated M\"ossbauer source allows one to temporally resolve the quantum beat \cite{Shvyd1999H} caused by the hyperfine splitting of the sample, which is so far only possible with SR \cite{Shenoy2006, Hastings1991, Burck1992}. 

In conclusion we have theoretically investigated a method using a $^{57}$Co X-ray source and a magnetically perturbed $^{57}$FeBO$_3$ crystal as a polarization transducer
to generate single hard X-ray pulses of tailored pulse durations in the range of 1 ns -100  ns. 
We expect that the present system can become an economic and complementary tool to other ultrashort hard X-ray sources for different purposes. Also, the adjustable pulse duration given by the present scheme offers flexible generations of time-bin qubits \cite{Olga2014}, arbitrary pulse trains and proper X-ray pulses for photon storage \cite{Kong2016}.

We thank S. M. Cavaletto, K. Heeg, J. Gunst, S. Bragin, A. P\'alffy, J. Evers, and B. Nickerson for carefully reading the manuscript and fruitful discussions.
G.-Y.~W. and W.-T.~L. are supported by the Ministry of Science and Technology, Taiwan (Grant No. MOST 105-2112-M-008-001-MY3). 
W.-T.~L. is also supported by the National Center for Theoretical Sciences, Taiwan.

\begin{widetext}

\appendix

\section{Optical-Bloch Equation For the Case of a fixed Magnetic Field}
The system of a fixed magnetic field under study is described by the Optical-Bloch equation (OBE) \cite{Scully2006,Liao2012a}
\begin{eqnarray}
 &  & \partial_{t}\hat{\rho}=\frac{1}{i\hbar}\left[\hat{H},\hat{\rho}\right]+\mathcal{L}\rho,\nonumber\\
 &  & \hat{H} = \hat{H}_i + \hat{H}_h ,\nonumber\\ 
 &  & \frac{1}{c}\partial_{t}\Omega_{R}+\partial_{y} \Omega_{R}=i\frac{6 \Gamma\alpha}{L}\left( a_{51}\rho_{51}+a_{62}\rho_{62}\right) ,\nonumber\\
 &  & \frac{1}{c}\partial_{t}\Omega_{H}+\partial_{y} \Omega_{H}=i\frac{6 \Gamma\alpha}{L}\left( a_{41}\rho_{41}+a_{52}\rho_{52}\right) ,\nonumber\\
 &  & \frac{1}{c}\partial_{t}\Omega_{L}+\partial_{y} \Omega_{L}=i\frac{6 \Gamma\alpha}{L}\left( a_{31}\rho_{31}+a_{42}\rho_{42}\right) ,
\end{eqnarray}
where $\alpha$ is the resonant thickness, $\hat{\rho}$ is the density matrix for the state vector
$\sum_{i=1}^{6}A_{i}\vert i\rangle$ of the 6-level $^{57}$Fe nuclei, and each coherence $\rho_{ij}=A_i A_j^*$; 
\begin{equation}
\hat{H}_{i}=-\frac{\hbar}{2}\left( 
\begin{array}{cccccc}
  0  & 0 & a_{13}\Omega_{L}^\ast & a_{14}\Omega_{H}^\ast & a_{15}\Omega_{R}^\ast & 0\\
  0  & 0 & 0 & a_{24}\Omega_{L}^\ast & a_{25}\Omega_{H}^\ast & a_{26}\Omega_{R}^\ast\\
  a_{31}\Omega_{L}  & 0 & -2\Delta_L & 0 & 0 & 0\\
  a_{41}\Omega_{H}  & a_{42}\Omega_{L} & 0 & -2\left( \Delta_L+\Delta_H\right)  & 0 & 0\\
  a_{51}\Omega_{R}  & a_{52}\Omega_{H} & 0 & 0 & -2\left( \Delta_H+\Delta_R\right) & 0\\
  0  & a_{62}\Omega_{R} & 0 & 0 & 0 & -2\Delta_R
\end{array}  
\right)
\end{equation}
is the interaction Hamiltonian
describing the nucleus-x-ray interaction with Rabi frequency $\Omega_R$  of right circularly polarized field, $\Omega_H$ of horizontally polarized field and $\Omega_L$ of the left circularly polarized field. 
$a_{31}=1$, $a_{41}=\sqrt{\frac{2}{3}}$, $a_{51}=\sqrt{\frac{1}{3}}$,  $a_{42}=\sqrt{\frac{1}{3}}$, $a_{52}=\sqrt{\frac{2}{3}}$ and $a_{62}=1$
are Clebsch-Gordan coefficients of corresponding transitions. All X-ray detunings are zero, where $\Delta_R$  is the detuning of right circularly polarized field, $\Delta_H$ of horizontally polarized field and $\Delta_L$ of the left circularly polarized field; 
\begin{equation}
\hat{H}_{h}=\hbar\left( 
\begin{array}{cccccc}
  -\delta_g  & 0 & 0 & 0 & 0 & 0\\
  0 &  \delta_g & 0 & 0 & 0 & 0\\
  0 & 0 & 3\delta_e & 0 & 0 & 0\\
  0 & 0 & 0 & \delta_e & 0 & 0\\
  0 & 0 & 0 & 0 & -\delta_e & 0\\
  0 & 0 & 0 & 0 & 0 & -3\delta_e
\end{array}  
\right)
\nonumber
\end{equation}
depicts the hyperfine splitting;
\begin{equation}
\mathcal{L}\rho=\frac{\Gamma}{2}\left( 
\begin{array}{cccccc}
  2\left( a_{31}^2\rho_{33}+a_{41}^2\rho_{44}+a_{51}^2\rho_{55}  \right)   & 0  & -\rho_{13} & -\rho_{14} & -\rho_{15} & -\rho_{16}\\
  0  & 2\left( a_{42}^2\rho_{44}+a_{52}^2\rho_{55}+a_{62}^2\rho_{66}  \right) & -\rho_{23} & -\rho_{24} & -\rho_{25} & -\rho_{26}\\
  -\rho_{31}  & -\rho_{32} & -2\rho_{33} & -2\rho_{34} & -2\rho_{35} & -2\rho_{36}\\
  -\rho_{41}  & -\rho_{42} & -2\rho_{43} & -2\rho_{44} & -2\rho_{46} & -2\rho_{46}\\
  -\rho_{51}  & -\rho_{52} & -2\rho_{53} & -2\rho_{54} & -2\rho_{55} & -2\rho_{56}\\
  -\rho_{61}  & -\rho_{62} & -2\rho_{63} & -2\rho_{64} & -2\rho_{65} & -2\rho_{66}
\end{array}  
\right)
\end{equation}
describes the radioactive decay of the excited states $\left|3\right\rangle $, $\left|4\right\rangle $, $\left|5\right\rangle $ and $\left|6\right\rangle $
characterized by decay rate $\Gamma=1/141$ GHz; 
$\alpha$ and $L$ the resonant thickness and the length of
the medium, respectively. 
The boundary condition gives the Rabi frequency of the incident x-ray field $\Omega_R \left( t,0\right)=\Omega_L \left( t,0\right) = e^{-\frac{\Gamma}{2} t}\times\frac{1}{2}\left[ 1+\tanh\left( 4\frac{t}{\tau}\right) \right] $, where the exponential decay depicts the radioactive decay from excited $^{57}$Fe nuclei, and hyperbolic tangent simulates the time gating with $\tau=5$ ns. Moreover, $\Omega_H \left( t,0\right)=0$ for the forward emission.
The initial conditions give the all initial x-ray fields are zero and initial population $\rho_{11}(0,y)=\rho_{22}(0,y)=1/2$,  $\rho_{ii}(0,y)=0$ for $i>2$ and zero coherences $\rho_{ij}(0,y)=0$ for $i\neq j$. 

\section{Optical-Bloch Equation For the Case of switched Magnetic Field}
For general cases of continuous and relatively slow magnetic switching,
it is convenient to fix the quantization axis along  z-axis and then rotate $\hat{H}_{h}$ at each $\theta$  back onto z-axis. We now derive the rotated master equation of the density matrix. The Schr\"odinger equation along the switched magnetic field are $i \hbar \frac{\partial}{\partial t} \vert \psi\rangle = \hat{H} \vert \psi\rangle$ and $-i \hbar \frac{\partial}{\partial t} \langle \psi\vert = \langle \psi\vert\hat{H}^\dagger$. The associated master equation is then give by 
\begin{equation}
\frac{\partial}{\partial t}\left( \vert \psi\rangle \langle \psi\vert\right)
= \left( \frac{\partial}{\partial t} \vert \psi\rangle \right) \langle \psi\vert
+ \vert \psi\rangle \left(  \frac{\partial}{\partial t}\langle \psi\vert\right). \nonumber
\end{equation}
By substituting Schr\"odinger equation on the right hand side, one gets
\begin{equation}
\frac{\partial}{\partial t}\left( \vert \psi\rangle \langle \psi\vert\right)
= \frac{1}{i\hbar}\left( \hat{H} \vert \psi\rangle \right) \langle \psi\vert
- \frac{1}{i\hbar} \vert \psi\rangle \left( \langle \psi\vert \hat{H}^\dagger \right), \nonumber
\end{equation}
which then becomes the typical master equation $\partial_{t}\hat{\rho}=\frac{1}{i\hbar}\left[\hat{H},\hat{\rho}\right]$. The same procedure of derivation can apply to the rotating case using the rotation operator $\hat{R}=e^{-i \frac{\hat{J}}{\hbar}\cdot\hat{n}\theta\left( t\right) }$ where $\hat{J}$ is the angular momentum operator, $\hat{n}$ the rotation axis and $\theta\left( t\right)$ is the time dependent switching angle between the z-axis and magnetic field at some moment $t$. Along the z-axis as the fixed quantization axis, we have
\begin{eqnarray}
\frac{\partial}{\partial t}\left( \hat{R}\vert \psi\rangle \langle \psi\vert\hat{R}^{\dagger}\right)
& = & \hat{R}\left( \frac{\partial}{\partial t} \vert \psi\rangle \right) \langle \psi\vert\hat{R}^{\dagger}
+ \hat{R}\vert \psi\rangle \left(  \frac{\partial}{\partial t}\langle \psi\vert\right)\hat{R}^{\dagger}
+ \left( \frac{\partial}{\partial t}\hat{R}\right) \vert \psi\rangle  \langle \psi\vert\hat{R}^{\dagger}
+ \hat{R}\vert \psi\rangle \langle \psi\vert\left(  \frac{\partial}{\partial t}\hat{R}^{\dagger}\right) \nonumber\\
& = & \frac{1}{i\hbar}\hat{R} \hat{H}\hat{R}^{\dagger} \hat{R}\vert \psi\rangle \langle \psi\vert\hat{R}^{\dagger}
- \frac{1}{i\hbar}\hat{R}\vert \psi\rangle \langle \psi\vert \hat{R}^{\dagger} \hat{R}\hat{H}^\dagger \hat{R}^{\dagger}
+ \left( \frac{\partial}{\partial t}\hat{R}\right) \vert \psi\rangle  \langle \psi\vert\hat{R}^{\dagger}
+ \hat{R}\vert \psi\rangle \langle \psi\vert\left(  \frac{\partial}{\partial t}\hat{R}^{\dagger}\right). \nonumber
\end{eqnarray}
When the magnetic firld is continuously switched, the $\hat{R}$ is time dependent, and the master equation turns into
\begin{eqnarray}
\partial_{t}\hat{\rho}^z
& = & \frac{1}{i\hbar}\left[\hat{H}^z,\hat{\rho}^z\right]
-i \frac{\hat{J}}{\hbar}\cdot\hat{n}\frac{\partial \theta}{\partial t}\hat{R} \vert \psi\rangle  \langle \psi\vert\hat{R}^{\dagger}
+ \frac{\partial \theta}{\partial t} \hat{R}\vert \psi\rangle \langle \psi\vert\hat{R}^{\dagger}i \frac{\hat{J}^{\dagger}}{\hbar}\cdot\hat{n}. \nonumber\\
& = & \frac{1}{i\hbar}\left[\hat{H}^z,\hat{\rho}^z\right]
+ \frac{\hat{J}\cdot\hat{n}}{i\hbar}\frac{\partial \theta}{\partial t}\hat{\rho}^z
- \frac{\partial \theta}{\partial t} \hat{\rho}^z\frac{\hat{J}^{\dagger}\cdot\hat{n}}{i\hbar}. \nonumber\\
& = & \frac{1}{i\hbar}\left[\hat{H}^z + \frac{\partial \theta}{\partial t}\hat{J}\cdot\hat{n},\hat{\rho}^z\right]. \nonumber
\end{eqnarray}
For simplicity we remove the z index of $\hat{\rho}^z$, the OBE of a time dependent rotating system becomes
\begin{eqnarray}
 &  & \partial_{t}\hat{\rho}=\frac{1}{i\hbar}\left[\hat{H},\hat{\rho}\right]+\mathcal{L}\rho,\nonumber\\
 &  & \hat{H} = \hat{H}_i + \hat{H}^z_h +\frac{\partial\theta}{\partial t}\hat{J}_y,
\end{eqnarray}
where $\hat{J}_y$ is the y-component of the angular momentum operator when  magnetic field is switched about y-axis \cite{Sakurai1994}. 
The $\theta$-independent $\hat{H}_i$ show the convenience of the scheme such that one can always use the same definition of X-ray polarizations.
The rotated hyperfine Hamiltonian reads
\begin{equation}
\hat{H}^z_{h}=\hat{R}\left( -\theta\right)\hat{H}_{h}\hat{R}^{\dagger}\left( -\theta\right)=\left( 
\begin{array}{cccccc}
  -\delta_g\cos\theta\left( t\right)  & -\delta_g\sin\theta\left( t\right) & 0 & 0 & 0 & 0\\
  -\delta_g\sin\theta\left( t\right) &  \delta_g\cos\theta\left( t\right) & 0 & 0 & 0 & 0\\
  0 & 0 & 3\delta_e\cos\theta\left( t\right) & \sqrt{3}\delta_e\sin\theta\left( t\right) & 0 & 0\\
  0 & 0 & \sqrt{3}\delta_e\sin\theta\left( t\right) & \delta_e\cos\theta\left( t\right) & 2\delta_e\sin\theta\left( t\right) & 0\\
  0 & 0 & 0 & 2\delta_e\sin\theta\left( t\right) & -\delta_e\cos\theta\left( t\right) & \sqrt{3}\delta_e\sin\theta\left( t\right)\\
  0 & 0 & 0 & 0 & \sqrt{3}\delta_e\sin\theta\left( t\right) & -3\delta_e\cos\theta\left( t\right)
\end{array}  
\right),
\nonumber
\end{equation}
where $\hat{R}\left( -\theta\right)$ is the rotation operator \cite{Sakurai1994}.

Due to the weak x-ray intensity of the $^{57}$Co source,   
in both cases we work in the perturbation regime, namely, $\vert\Omega_R\vert \ll \Gamma$, $\vert\Omega_H\vert \ll \Gamma$ and $\vert\Omega_L\vert \ll \Gamma$ such that only terms of $\rho_{i1}$ and $\rho_{j2}$, where $i > 2$ and $j > 2$, are used in the calculation. $\alpha=30$, $\delta_g=12.63\Gamma$ and $\delta_e=7.37\Gamma$ are used for all figures in the main text.

Because the X-ray propagation time, namely, $L/c$, is on the order of 0.1 ps which is much shorter than the nanosecond time scale of interest, this allows for the typical procedure to neglect the temporal derivative terms in the wave equations. The OBE  can then be numerically solved by Runge Kutta 4th order method (RK4) \cite{Press1996} with $L= 10 \mu$m, grid spacings $\Delta y =0.1 \mu$m and  $\Delta t = 3\times 10^{-3}$ ns. All solutions are double checked with NDSolve of Mathematica 11 for complete ODE including temporal derivative terms in the wave equations. No significant deviation is observed, which confirms the convergence of the solution.

\section{Estimation of Photon counting rates}
We estimate the vertically polarized photon counting rate $\chi_1$ at the first crystal by following formula in the spherical coordinate
\begin{eqnarray}
\chi_1 &=& A \frac{\Theta}{4\pi} P_V , \\
\Theta  &=& \int_0^{2\pi}\int_0^{\tan^{-1}\left(\frac{w}{2 D} \right) }\sin\vartheta d\vartheta d\phi .
\end{eqnarray}
Here 
the activity of 200 mCi $^{57}$Co source is $A=7.4\times 10^{9}$ counts/s.
When $\gamma$-decay equally happens in total $4\pi$ solid angle, one has to calculate the fraction of photons emitted in the forward direction within a small solid angle $\Theta$. 
As depicted in Fig.~\ref{figs1}, assuming the distance between a radioisotope source and the first crystal is $D=20$ cm and 
the width of the first $^{57}$FeBO$_3$ crystal is $w=$ 5 mm,  these parameters result in a solid angle of $5 \times 10^{-4}$ sr in the forward direction.
In order to calculate the probability $P_V$ of the emission of vertically polarized photons by a $^{57}$Co source,
one can choose the x-axis as the quantization axis and then look at the $\Delta m = 0$ transitions. The analysis of Clebsch-Gordan coefficients shows $P_V=\frac{a_{41}^2+a_{52}^2}{a_{41}^2+a_{52}^2+a_{51}^2+a_{62}^2+a_{31}^2+a_{42}^2}=\frac{1}{3}$. 
With above parameters, $\chi_1=96343$ counts/s is obtained.
The production rate $\chi_2$ of horizontally polarized photons  at the second crystal can be estimated by
\begin{equation}\label{eqs5}
\chi_2=\chi_1  \frac{\int_0^\infty \vert \Omega_H\left( t,L\right) \vert^2 dt}{2 \int_0^\infty e^{-\Gamma t} dt}.
\end{equation}
The denominator represents the total temporal area of the incident VPX pulse, and the numerator denotes that of generated HPX pulse.
The area ratio in Eq.~(\ref{eqs5}) gives the conversion efficiency from vertically to horizontally polarized x-rays. For three single pulses in Fig.~3(c), the area ratios are respectively 0.0045, 0.069 and 0.019, and so the HPX photon production rates are in turn estimated to be 430/s, 6600/s and 1800/s.  The area ratio of 1.3 ns pulse in Fig.~3(c) is tripled as 0.013 by using a thicker $^{57}$FeBO$_3$ crystal of $\alpha=100$, $\delta_e=31\Gamma$ and $\delta_g=53\Gamma$. The covered horizontally polarized x-rays is demonstrated in Fig.~\ref{figs2}(b).
\begin{figure}[t]
\vspace{-0.4cm}
 \includegraphics[width=0.45\textwidth]{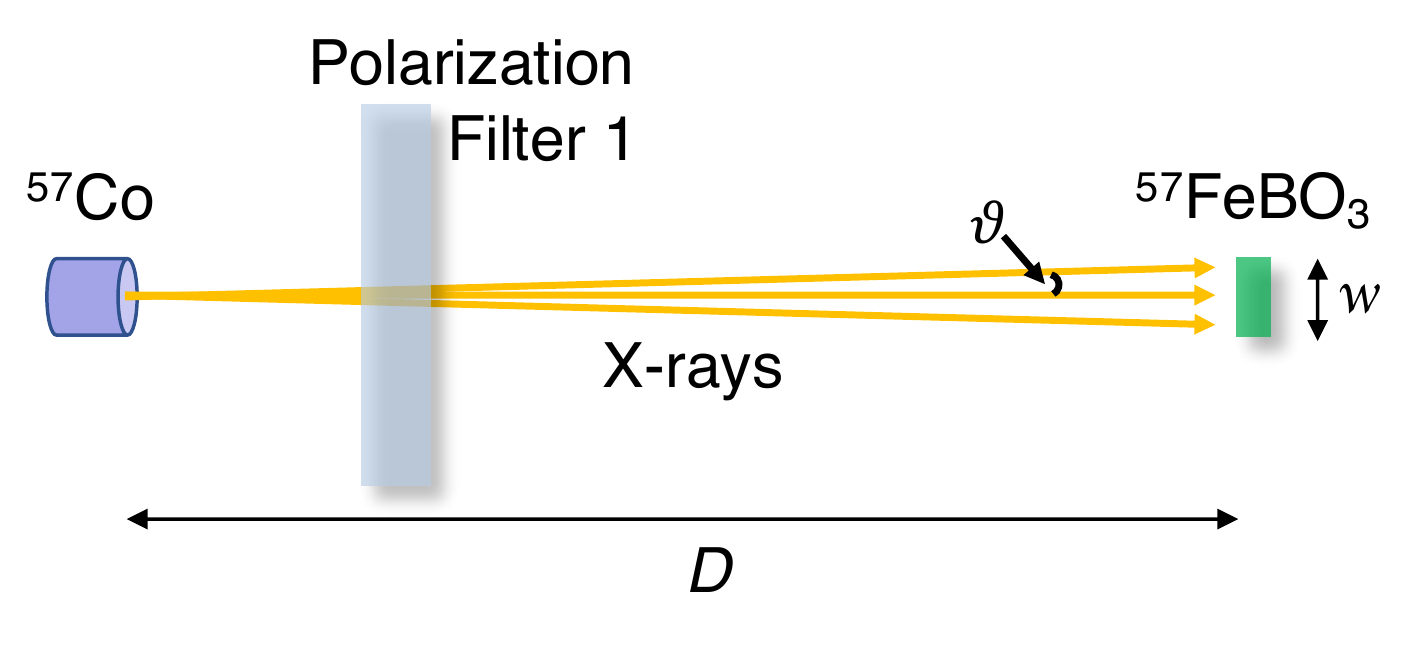}
  \caption{\label{figs1}
The illustration for the estimation of photon counting rates. The distance $D$ between a radioisotope source and the $^{57}$FeBO$_3$ crystal  and the crystal width $w$ are used to calculate the solid angle. $\vartheta$ denotes the polar angle in the spherical coordinate.
  }
\end{figure}
\begin{figure}[b]
\vspace{-0.4cm}
 \includegraphics[width=0.45\textwidth]{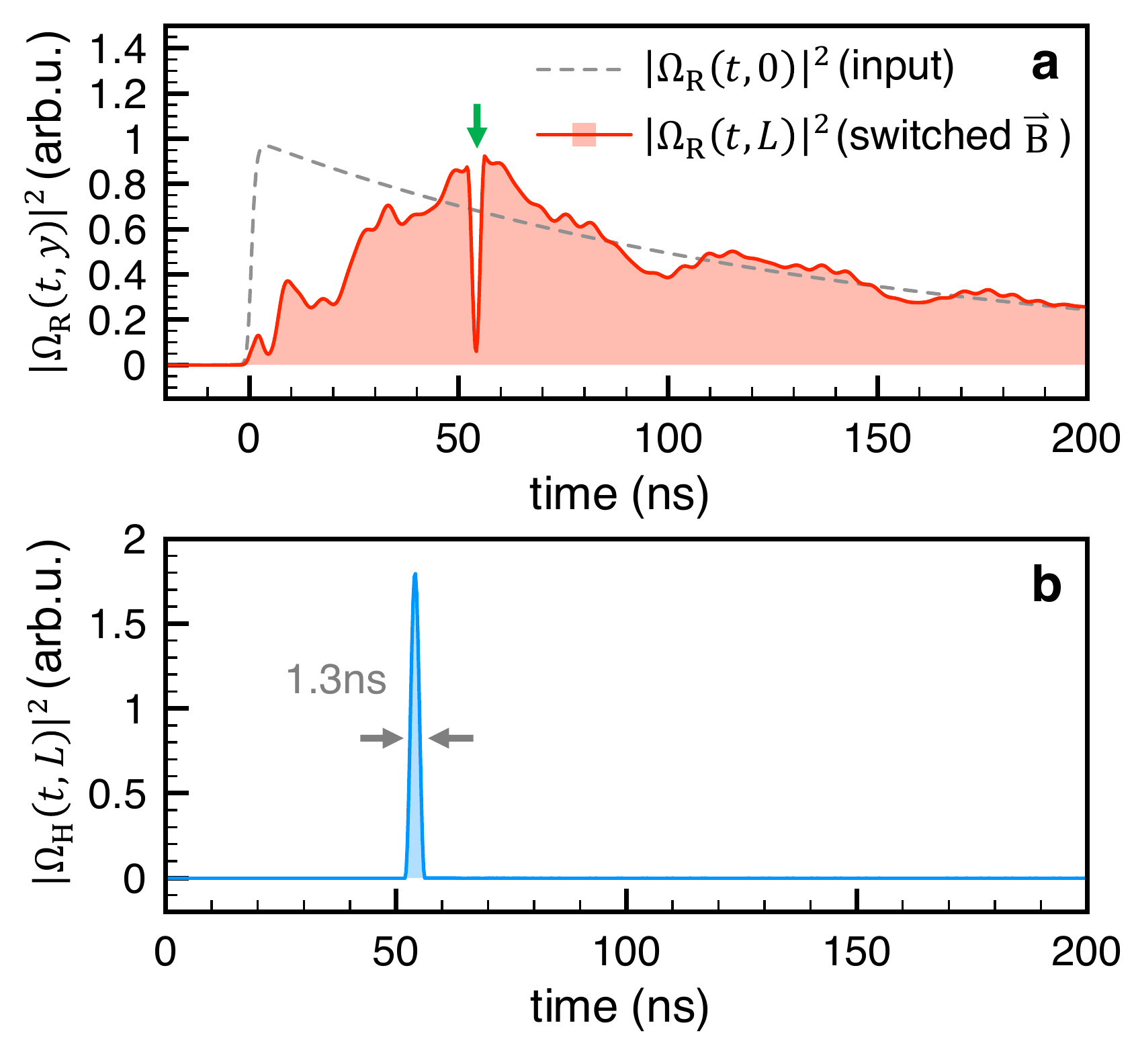}
  \caption{\label{figs2}
Comparing to Fig.~3, photon counting rates of horizontally polarized x-ray  at the second target are  tripled by using a thicker $^{57}$FeBO$_3$ crystal of $\alpha=100$, $\delta_e=31\Gamma$ and $\delta_g=53\Gamma$. 
  }
\end{figure}
\end{widetext}
\clearpage

%
\bibliographystyle{apsrev}
\bibliography{NFSBS2A}

\end{document}